\documentclass[preprintnumbers, prd, twocolumn, showpacs, floatfix, preprintnumbers, letterpaper, superscriptaddress,nofootinbib]{revtex4}
\usepackage{graphicx,epsfig}
\usepackage{amsmath}
\usepackage {amssymb}

\usepackage{relsize}
\usepackage{relsize}

\newcommand{\be}{\begin{equation}}
\newcommand{\ee}{\end{equation}}
\newcommand{\bea}{\begin{eqnarray}}
\newcommand{\eea}{\end{eqnarray}}

\begin{document}

\title{Nonminimal torsion-matter coupling extension of $f(T)$ gravity}

\author{Tiberiu Harko}
\email{t.harko@ucl.ac.uk}
\affiliation{Department of Mathematics, University College London, Gower
Street, London
WC1E 6BT, United Kingdom}

\author{Francisco S. N. Lobo}
\email{flobo@cii.fc.ul.pt}
\affiliation{Centro de Astronomia e Astrof\'{\i}sica da Universidade de
Lisboa, Campo
Grande, Edific\'{\i}o C8, 1749-016 Lisboa, Portugal}

\author{G. Otalora}
\email{gotalora@fisica.ufjf.br}
\affiliation{Departamento de F\'{\i}sica, ICE, Universidade Federal de Juiz
de Fora, Caixa Postal 36036-330, Minas Gerais, Brazil}

\author{Emmanuel N. Saridakis}
\email{Emmanuel\_Saridakis@baylor.edu}
\affiliation{Physics Division,
National Technical University of Athens, 15780 Zografou Campus,
Athens, Greece} \affiliation{Instituto de F\'{\i}sica, Pontificia
Universidad de Cat\'olica de Valpara\'{\i}so, Casilla 4950,
Valpara\'{\i}so, Chile}

\begin{abstract}
We construct an extension of $f(T)$ gravity with the inclusion of a
non-minimal torsion-matter coupling in the action. The resulting theory is a
novel gravitational modification, since it is different from both $f(T)$
gravity, as well as from the nonminimal curvature-matter-coupled theory. The
cosmological application of this new theory proves to be very interesting. In
particular, we obtain an effective dark energy sector whose equation-of-state
parameter can be quintessence- or phantom-like, or exhibit the phantom-divide
crossing, while for a large range of the model parameters the universe
results in a de Sitter, dark-energy-dominated, accelerating phase.
Additionally, we can obtain early-time inflationary solutions too, and thus
provide a unified description of the cosmological history.
\end{abstract}


\maketitle

\section{Introduction}\label{Introduction}

The recent observational advances in cosmology have provided a large amount
of high-precision cosmological data, which has posed new challenges   for
the understanding of the basic physical properties of the Universe, and of
the gravitational interaction that dominates its dynamics and evolution. The
observation of the accelerated expansion of the Universe
\cite{Riess1} has raised the fundamental issue of the cause
of this acceleration, which is usually attributed to a mysterious and yet not
directly detected dominant component of the Universe, called dark energy
\cite{Copeland:2006wr}.
In this context, the recently released Planck satellite data of the 2.7 degree Cosmic
Microwave Background  (CMB) full sky survey \cite{Planckresults} have generally
confirmed the standard $\Lambda $ Cold Dark Matter ($\Lambda $CDM)
cosmological model. On the other hand, the measurement of the tensor modes
from large angle CMB B-mode polarisation by BICEP2 \cite{BICEP2}, implying a
tensor-to-scalar ratio $r = 0.2 ^{+0.07}_{-0.05}$, has provided a very
convincing evidence for the inflationary scenario, since  the generation of
gravitational wave fluctuations is a generic prediction of the early de
Sitter exponential expansion. However, the BICEP2 result is in tension with
Planck limits on standard inflationary models \cite{Mar}, and thus
alternative
explanations may be required.  In principle, magnetic fields generated during
inflation can produce the required B-mode, for a suitable range of energy
scales of inflation \cite{Mar}. Moreover, the existence of the fluctuations
of cosmological birefringence can give rise to CMB B-mode polarization that
fits BICEP2 data with $r<0.11$,  and no running of the scalar spectral index
\cite{Lee}.  

The above major observational advances require some good theoretical
explanations, with the role of giving a firm foundation to cosmology, and the
underlying theory of gravity. However, up to now, no convincing theoretical
model, supported by observational evidence that could clearly explain the
nature of dark energy, has been proposed. Moreover, not only the recent
accelerated expansion of the Universe,  but also observations at the galactic
(galaxy rotation curves)  and extra-galactic scale (virial mass
discrepancy in galaxy clusters) \cite{Str} suggest the existence of another
mysterious and yet undetected major component of the Universe, the so-called
dark matter. From all these observations one can conclude  that the
standard general relativistic gravitational field equations, obtained from 
the classic Einstein-Hilbert
action $S=\int{\left(R/2+L_m\right)\sqrt{-g}d^4x}$, where $R$ is the scalar
curvature, and $L_m$ is the matter Lagrangian density,  in which matter is
minimally coupled to the geometry, cannot give an appropriate quantitative
description of the Universe at astrophysical scales going beyond the boundary
of the Solar System. To explain dark energy and dark matter in a
cosmological context requires the {\it ad hoc} introduction of the dark
matter and dark energy components into the  total energy-momentum tensor of the
Universe, in addition to the ordinary baryonic matter.

From a historical point of view, in going beyond the Einstein-Hilbert action,
the first steps were taken in the direction of generalizing the geometric
part of the standard gravitational action.  An extension of the
Einstein-Hilbert action, in which the Ricci scalar invariant $R$ is
substituted with an arbitrary function of the scalar invariant, $f(R)$, has
been extensively explored in the literature \cite{rev}. Such a modification
of the gravitational action can explain the late acceleration of the
Universe, and may also provide  a geometric explanation for dark matter,
which can be described as a manifestation of  geometry itself  \cite{Boehm}.
Furthermore, quadratic Lagrangians, constructed from second order curvature
invariants such as $R^{2}$, $R_{\mu \nu
}R^{\mu\nu }$, $R_{\alpha \beta \mu \nu }R^{\alpha \beta \mu \nu }$,
$\varepsilon ^{\alpha \beta \mu \nu }R_{\alpha \beta \gamma \delta
}R_{\mu \nu }^{\gamma \delta }$, $C_{\alpha \beta \mu \nu}C^{\alpha \beta \mu
\nu}$, etc., have also been considered as candidates for a more general
gravitational actions \cite{Lobo:2008sg}, which can successfully explain dark
matter and the late-time cosmic acceleration. Alternatively, the interest for
extra-dimensions, which goes back to the unified field theory of Kaluza and
Klein, led to the development of the braneworld models \cite{Maartens1}. In
braneworld models, gravitational effects due to the
extra dimensions dominate at high energies, but important new effects, which
can successfully explain both dark energy and dark matter, also appear at low
energies.

Most of the modifications of the Einstein-Hilbert Lagrangian involve a 
change in the geometric part of the action only,  and assume that the matter
Lagrangian plays a subordinate and passive role, which is implemented by
the minimal coupling of matter to geometry. However, a general theoretical
principle forbidding an arbitrary coupling between matter and geometry does
not exist {\it a priori}. If theoretical models, in which matter is
considered on an equal footing with geometry, are allowed, gravitational
theories with many interesting and novel features can be constructed.

A theory with an explicit coupling between an arbitrary function of the
scalar curvature and the Lagrangian density of matter was proposed in
\cite{Bertolami:2007gv}. The gravitational action of the latter model is of the form
$S=\int{\left\{f_1(R)+\left[1+\lambda
f_2(R)\right]L_m\right\}\sqrt{-g}d^4x}$. In these models an extra force
acting on massive test particles arises, and the motion is no longer
geodesic. Moreover, in this framework, one can also explain dark matter 
\cite{Bertolami:2009ic}.
The early ``linear''  geometry-matter coupling \cite{Bertolami:2007gv} was
extended in \cite{Harko:2008qz} and a maximal extension of the Einstein-Hilbert
action with geometry-matter coupling, of the form $S=\int d^{4}x
\sqrt{-g}f\left(R,L_m\right)$ was considered in \cite{Harko:2010mv}. An
alternative model to $f(R,L_m)$ gravity is the $f(R,\mathcal{T})$
theory \cite{Harko:2011kv}, where $\mathcal{T}$ is the trace of the matter
energy-momentum tensor
$T_{\mu\nu}$, and the corresponding action given by
$S=\int{\left[f\left(R,\mathcal{T}\right)/2+
L_m\right]\sqrt{-g}d^4x}$. The dependence of the gravitational action on
$\mathcal{T}$ may be
due to the presence of  quantum effects (conformal anomaly), or of some
exotic imperfect fluids.  When the trace of the energy-momentum tensor
$\mathcal{T}$  is zero, $\mathcal{T}=0$, which is the case for the
electromagnetic
radiation, the field equations of $f(R,T)$ theory reduce to those of 
$f(R)$ gravity.

However, the $f(R,L_m)$ or $f(R,\mathcal{T})$ gravitational models are not
the most general
Lagrangians with nonminimal geometry-matter couplings. One could further
obtain interesting gravity models by introducing a term of the form
$R_{\mu\nu}T^{\mu\nu}$ into the Lagrangian \cite{fRT,Od}. Such couplings
appear in Einstein-Born-Infeld theories \cite{deser}, when one expands the
square root in the Lagrangian. The presence of the $R_{\mu\nu}T^{\mu\nu}$
coupling term has the advantage of entailing a nonminimal coupling of
geometry to the electromagnetic field.

All the above gravitational modifications are based on the Einstein-Hilbert
action, namely on the curvature description of gravity. However, an
interesting and rich class of modified gravity can arise if one modifies
the action of the equivalent torsional formulation of General Relativity.
As it is well known, Einstein also constructed the ``Teleparallel Equivalent
of General Relativity'' (TEGR)
\cite{Unzicker:2005in,TEGR,Hayashi:1979qx,JGPereira,Maluf:2013gaa}, replacing
the torsion-less Levi-Civita connection by the curvature-less
Weitzenb{\"{o}}ck one, and using the vierbein instead of the metric as the
fundamental field. In this formulation, instead of the curvature (Riemann)
tensor one has the torsion tensor, and the Lagrangian of the theory, namely
the torsion scalar $T$, is constructed by contractions of the torsion tensor.
Thus, if one desires to modify gravity in this formulation, the simplest
thing is to extend $T$ to an arbitrary function$f(T)$
\cite{Ferraro:2006jd,Linder:2010py}. An interesting aspect of this extension 
is that
although TEGR coincides with General Relativity at the level of equations,
$f(T)$ is different than $f(R)$, that is they belong to different
modification classes. Additionally, although  in $f(R)$ theory the
field equations are fourth order, in $f(T)$ gravity they are  second
order, which is a great advantage. $f(T)$ gravity models have been
extensively applied to cosmology, and amongst other applications it is able to explain
the late-time accelerating expansion of the Universe without the need
for dark energy
\cite{Linder:2010py,Chen:2010va,Bengochea001}. Furthermore, following these
lines, and
inspired by the higher-curvature modifications of General Relativity, one can
construct gravitational modifications based on higher-order torsion
invariants, such is the $f(T,T_G)$ gravity \cite{Kofinas:2014owa}, which also
proves to have interesting cosmological implications.

Another gravitational modification based on the teleparallel formulation is
the generalization of TEGR to the case of a Weyl-Cartan space-time, in which
the Weitzenb\"{o}ck condition of the vanishing of the curvature is also
imposed (Weyl-Cartan-Weitzenb\"{o}ck (WCW) gravity), with the addition of a
kinetic term for the torsion in the gravitational action \cite{WC1}. In
this framework the late-time acceleration of the Universe can be naturally obtained,
determined by the intrinsic geometry of the space-time. A further extension
of the WCW gravity, in which the Weitzenb\"{o}ck condition in a Weyl-Cartan
geometry is inserted into the gravitational action via a Lagrange multiplier,
was analyzed in \cite{WC2}. In the weak field limit the gravitational
potential explicitly depends on the Lagrange multiplier and on the Weyl
vector, leading to an interesting cosmological behavior.

In the work, we are interested in  proposing a novel gravitational
modification based on the torsional formulation, by allowing the
possibility of a nonminimal torsion-matter coupling in the gravitational
action. In particular, for the torsion-matter coupling we adopt the
``linear'' model introduced in the case of $f(R)$ gravity in
\cite{Bertolami:2007gv}. Hence, the gravitational field can be described in
terms of two arbitrary functions of the torsion scalar $T$, namely $f_1(T)$
and $f_2(T)$, with the function $f_2(T)$ linearly coupled to the matter
Lagrangian. This new coupling induces a supplementary term $\left[1+\lambda
f_2(T)\right]L_m$ in the standard $f(T)$ action, with $\lambda$ an
arbitrary coupling constant. When $\lambda =0$, the model reduces to the
usual $f(T)$ gravity. We investigate in detail the cosmological implications
of the torsion-matter coupling for two particular choices of the functions
$f_1(T)$ and $f_2(T)$. For both choices the Universe evolution is in
agreement with the observed behavior, and moreover it ends in a de Sitter
type vacuum state, with zero matter energy density. The details of the
transition depend on the numerical values of the free parameters that appear
in the functions $f_1(T)$ and $f_2(T)$.

The paper is organized as follows. In Section~\ref{fTmodel} we briefly
describe the basics of the $f(T)$ gravity model. The field equations of the
$f(T)$ theory with linear nonminimal torsion-matter coupling are obtained in
Section~\ref{matter}. The cosmological implications of the theory are
analyzed in Section~\ref{cosm}. Finally, we conclude and discuss our
results in Section~\ref{Concl}.

\section{$f(T)$ gravity and Cosmology}\label{fTmodel}

In this Section, we briefly review the $f(T)$ gravitational paradigm. We use
the notation where Greek indices run over the coordinate space-time and
Latin indices run over the tangent space-time. As we mentioned in the
Introduction, the dynamical variables are the vierbein fields
${\mathbf{e}_A(x^\mu)}$, which at each point $x^\mu$ of the manifold form an
orthonormal basis for the tangent space, that is $\mathbf{e}
_A\cdot%
\mathbf{e}_B=\eta_{AB}$, with $\eta_{AB}={\rm diag} (1,-1,-1,-1)$.
Additionally, they can be expressed in terms of the components $e_A^\mu$ in
the coordinate basis as $\mathbf{e}_A=e^\mu_A\partial_\mu$.
Hence, the metric is obtained from the
dual vierbein through
\begin{equation}  \label{metrdef}
g_{\mu\nu}(x)=\eta_{AB}\, e^A_\mu (x)\, e^B_\nu (x).
\end{equation}
In this formulation, instead of the Levi-Civita connection one uses the
Weitzenb\"{o}ck one:
$\overset{\mathbf{w}}{\Gamma}^\lambda_{\nu\mu}\equiv e^\lambda_A\:
\partial_\mu
e^A_\nu$ \cite{Weitzenb23}, and thus instead of curvature we acquire the 
 torsion tensor  
\begin{equation}
\label{torsion2}
{T}^\lambda_{\:\mu\nu}=\overset{\mathbf{w}}{\Gamma}^\lambda_{
\nu\mu}-%
\overset{\mathbf{w}}{\Gamma}^\lambda_{\mu\nu}
=e^\lambda_A\:(\partial_\mu
e^A_\nu-\partial_\nu e^A_\mu).
\end{equation}
It proves convenient to define the contorsion tensor
$K^{\mu\nu}{}_{\rho}\equiv-\frac{1}{2}\Big(T^{\mu\nu}{}_{\rho}
-T^{\nu\mu}{}_{\rho}-T_{\rho}{}^{\mu\nu}\Big)$, as well as the tensor
$
S_{\rho}{}^{\mu\nu}\equiv\frac{1}{2}\Big(K^{\mu\nu}{}_{\rho}
+\delta^\mu_\rho
\:T^{\alpha\nu}{}_{\alpha}-\delta^\nu_\rho\:
T^{\alpha\mu}{}_{\alpha}\Big)$. Using these one can write down the 
teleparallel Lagrangian (torsion scalar)
\cite{TEGR,Hayashi:1979qx,JGPereira,Maluf:2013gaa,
Maluf:1994ji} 
\begin{equation}
\label{torsionscalar}
T\equiv\frac{1}{4}
T^{\rho \mu \nu}
T_{\rho \mu \nu}
+\frac{1}{2}T^{\rho \mu \nu }T_{\nu \mu\rho}
-T_{\rho \mu}{}^{\rho }T^{\nu\mu}{}_{\nu},
\end{equation}
which used in the action and varied in terms of the vierbeins gives
rise to the same equations with General Relativity. That is why such a
theory is called ``Teleparallel Equivalent of General Relativity'' (TEGR).

One can be based on the above torsional formulation of General Relativity, in
order to construct classes of modified gravity. The simplest one is to extend
$T$ to a function $T+f(T)$, that is writing an action of the
form\footnote{An alternative simple extension of TEGR is  to allow for a
nonminimal scalar-torsion coupling, as in \cite{Geng:2011aj}.}
\begin{eqnarray}
\label{action00}
S = \frac{1}{16\pi G}\int d^4x e \left[T+f(T)\right],
\end{eqnarray}
where $e = \text{det}(e_{\mu}^A) = \sqrt{-g}$, $G$ is the gravitational
constant, and we have used units where the speed of light is $c=1$. Note
that TEGR and thus General Relativity is restored when $f(T)=0$.  Moreover,
we stress that although TEGR coincides with General Relativity at
the level of equations, $f(T)$ is different than $f(R)$. 

Let us now proceed to the cosmological application of $f(T)$ gravity.
Introducing additionally the matter sector  the total
action becomes
\begin{eqnarray}
\label{actionaa}
S= \frac{1}{16\pi G }\int d^4x e
\left[T+f(T)+L_m\right],
\end{eqnarray}
where the matter Lagrangian is assumed to correspond
to a perfect fluid with energy density $\rho_m$ and pressure $p_m$ (for
simplicity we neglect the radiation sector, although its
inclusion is straightforward). Varying the action (\ref{actionaa}) with
respect to the vierbeins we obtain the field equations
\begin{eqnarray}\label{eom}
&&\left(1+f'\right)\,\left[e^{-1}\partial_{\mu}(ee_A^{\rho}S_{\rho}{}^{\nu\mu
}) -e_{A}^{\lambda}T^{\rho}{}_{\mu\lambda}S_{\rho}{}^{\mu\nu}\right]
\nonumber\\
&&  \hspace{-0.5cm}
 +
e_A^{\rho}S_{\rho}{}^{\nu\mu}\partial_{\mu}{T} f''+\frac{1}{
4} e_ {A}^{\nu}[T+f] = 4\pi Ge_{A}^{\rho}\overset {\mathbf{em}}T_{\rho}{}^{\nu},
\end{eqnarray}
where $f'=\partial f/\partial T$, $f''=\partial^{2} f/\partial T^{2}$,
and $\overset{\mathbf{em}}{T}_{\rho}{}^{\nu}$  denotes the usual
energy-momentum tensor.

Proceeding forward, we impose the standard homogeneous and isotropic
geometry, that is we consider
\begin{equation}
\label{weproudlyuse}
e_{\mu}^A={\rm
diag}(1,a(t),a(t),a(t)),
\end{equation}
which corresponds to a flat Friedmann-Robertson-Walker (FRW) universe with
metric
\begin{equation}
ds^2= dt^2-a^2(t)\,\delta_{ij} dx^i dx^j,
\end{equation}
where $a(t)$ is the scale factor.

In summary, inserting the vierbein ansantz (\ref{weproudlyuse}) into the
equations of motion (\ref{eom}) we extract the modified Friedmann
equations as
\begin{eqnarray}\label{background1}
&&H^2= \frac{8\pi G}{3}\rho_m
-\frac{f}{6}-2H^2f'\\
\label{background2}
&&\dot{H}=-\frac{4\pi G(\rho_m+p_m)}{1+f'-12H^2f''},
\end{eqnarray}
with
$H\equiv\dot{a}/a$  the Hubble parameter, and  dots denoting derivatives with
respect to $t$. Note that we have also used the relation
\begin{eqnarray}
\label{TH2}
T=-6H^2,
\end{eqnarray}
which arises immediately for an FRW geometry using Eq. (\ref{torsionscalar}).

\section{ $f(T)$ gravity with nonminimal torsion-matter coupling}\label{matter}

Having presented the $f(T)$ modified gravity in the previous section, in
this section we extend it, allowing for a nonminimal coupling between the
torsion scalar and the matter Lagrangian. In particular, we consider the
action
\begin{equation}
S= \frac{1}{16\pi G}\,\int
d^{4}x\,e\,\left\{T+f_{1}(T)+\left[1+\lambda\,f_{2}
(T)\right]\,L_{m}\right\},
\label{1}
\end{equation}
where $f_{i}(T)$ (with $i=1,2$) are arbitrary functions of the
torsion scalar $T$ and $\lambda$ is a coupling constant with units of ${\rm
mass}^{-2}$. Varying the action with respect to the tetrad $e^{A}_{\rho}$
yields the field
equations
\begin{eqnarray}
&&\left(1+f_{1}'+\lambda f_{2}' L_{m}\right) \left[e^{-1}
\partial_{\mu}{(e e^{\alpha}_{A} S_{\alpha}{}^{\rho \mu})}-e^{\alpha}_{A}
T^{\mu}{}_{\nu \alpha} S_{\mu}{}^{\nu\rho}\right]\nonumber\\
&&\ \ \ \ \ 
+\left(f_{1}''+ \lambda f_{2}'' L_{m}
\right)  \partial_{\mu}{T} e^{\alpha}_{A} S_{\alpha}{}^{\rho\mu}+
e_{A}^{\rho} \left(\frac{f_{1}+T}{4}\right)
\nonumber\\
&&\ \ \ \ \ 
-\frac{1}{4} \lambda f_{2}' \,
\partial_{\mu}{T} e^{\alpha}_{A} \overset{\mathbf{em}}{S}_{\alpha}{}^{\rho \mu}
+ \lambda f_{2}'\, e^{\alpha}_{A} S_{\alpha}{}^{\rho\mu} \, \partial_{\mu}{L_{m}}
\nonumber\\
&&\ \ \ \ \ 
=4\pi G \left(1+\lambda f_{2}\right) e^{\alpha}_{A}
\overset{\mathbf{em}}{T}_{\alpha}{}^{\rho},
\label{geneoms}
\end{eqnarray}
where we have defined
\begin{equation}
\overset{\mathbf{em}}{S}_{A}{}^{\rho
\mu}=\frac{\partial{L_{m}}}{\partial{\partial_{\mu}{e^{A}_{\rho}}}},
\label{Stilde}
\end{equation} 
and the prime denotes differentiation with respect to the torsion
scalar. As expected Eq. (\ref{geneoms}) reduces to Eq. (\ref{eom}) when
$\lambda=0$.

Since the Lagrangian density of a perfect fluid is the energy scalar,
representing the energy in a local rest frame for the fluid, a possible
``natural choice'' for the matter Lagrangian density is
$L_{m}/(16 \pi G)=-\rho_{m}$ \cite{GroenHervik,BPHL}.
In this case, we have $\overset{\mathbf{em}}{S}_{A}{}^{\rho \mu}=0$, and also the usual form
of the energy momentum tensor for the perfect fluid
$\overset{\mathbf{em}}{T}_{\mu\nu}=(\rho_{m}+p_{m}) u_{\mu} u_{\nu}-p_{m} g_{\mu
\nu}$.

In summary, inserting the flat FRW vierbein choice (\ref{weproudlyuse}) and
the above  matter Lagrangian density, into the field equations
\eqref{geneoms}, we obtain the modified Friedmann equations
\begin{equation}\label{H0}
 H^{2}=\frac{8\pi G}{3} \left[1+\lambda
 \left(f_{2}+12 H^{2} f_{2}' \right)\right] \rho_m-\frac{1}{6} \left(f_{1}+12
H^{2} f_{1}'\right),
\end{equation}
\begin{equation}\label{H00}
 \dot{H}=-\frac{4\pi G\left( \rho _{m}+p_{m}\right) \left[ 1+\lambda
\left(
f_{2}+12H^{2}f_{2}^{\prime }\right) \right] }{1+f_{1}^{\prime
}-12H^{2}f_{1}^{\prime \prime }-16 \pi G \lambda \rho _{m}\left( f_{2}^{\prime
}-12H^{2}f_{2}^{\prime \prime }\right) }.
\end{equation}
In the limit $\lambda =0$, $f_1(T)\equiv f(T)$, and $f_2(T)\equiv 0$, 
Eqs.~(\ref{H0}) and (\ref{H00}) reduce to Eqs.~(\ref{background1}) and
(\ref{background2}), respectively.
The generalized Friedmann equations  can be rewritten  as
 \begin{eqnarray}
  3H^2&=& 8\pi G\left(\rho_{DE}+\rho_m  \right), \label{Fr1} \\
  2\dot{H}+3H^2& =&-8\pi G\left(p_{DE}+p_m\right), \label{Fr2}
\end{eqnarray}
where the effective energy density and effective pressure of the dark energy 
sector are defined as
\begin{equation}
\label{rhode}
 \rho_{DE}:=-\frac{1}{16 \pi G} \left(f_{1}+12 H^{2}
f_{1}'\right)+ \lambda \rho_m\left(f_{2}+12 H^{2} f_{2}' \right) ,
\end{equation}
\begin{eqnarray}\label{pde}
&&p_{DE}:= \left(\rho _m+p_m\right)\times \nonumber\\
&&\left[\frac{  1+\lambda \left(
f_{2}+12H^{2}f_{2}^{\prime }\right)  }{1+f_{1}^{\prime
}-12H^{2}f_{1}^{\prime \prime }-16\pi G\lambda \rho _{m}\left( f_{2}^{\prime
}-12H^{2}f_{2}^{\prime \prime }\right) }-1\right]\nonumber\\
&&+\frac{1}{16 \pi G} \left(f_{1}+12 H^{2}
f_{1}'\right)- \lambda \rho_m\left(f_{2}+12 H^{2} f_{2}' \right).
\end{eqnarray}
Furthermore, we can define the dark-energy equation-of-state parameter
in the standard form
\begin{eqnarray}
w_{DE}:= \frac{p_{DE}}{\rho_{DE}}.
\label{wDE}
\end{eqnarray} 
One can easily verify that the above affective dark energy density
and pressure satisfy the usual evolution equation
\begin{eqnarray}\label{cons}
 \dot{\rho}_{DE}+\dot{\rho} _m +3H\left(\rho_{DE}+\rho
_m+p_{DE}+p_m\right)=0.
\end{eqnarray}
Finally, we can introduce the deceleration parameter $q$, given by
\be
q=\frac{d}{dt}\left(\frac{1}{H}\right)-1,
\ee
whose sign indicates the decelerating/accelerating nature of the cosmological
expansion. Cosmological models with $q<0$ are accelerating, while those
having $q>0$ experience a decelerating evolution.

\section{Cosmological implications}\label{cosm}

Since we have extracted the basic background equations of motion of the
$f(T)$ gravity model with a nonminimal matter-torsion coupling, we are now
able to investigate its phenomenological implications. Due to the relation
(\ref{TH2}), for convenience in the following we will change the
$T$-dependence to the $H$-dependence in the involved expressions, so that
$f_1(T)\equiv f_1(H)$, and $f_2(T)\equiv f_2(H)$. For the derivatives of the
functions $f_1(T)$ and $f_2(T)$ we obtain $f_i'(H)\equiv \left .f_i'(T)
\right |_{T\rightarrow -6H^2}$, and $f_i''(H)\equiv \left .f_i''(T) \right
|_{T\rightarrow -6H^2}$, $i=1,2$, respectively. Finally, in the following we
fully adopt the natural system of units by taking $8\pi G=c=1$.

The basic cosmological equations describing the time evolution of the
nonminimally-coupled $f(T)$ gravity are given by Eqs.~(\ref{H0}) and
(\ref{H00}). From Eq.~(\ref{H0}) we can express the matter
density as
\be\label{c1}
\rho _m(t)=\frac{3H^2+\left[f_1(H)+12H^2\left. f_1'(T)\right  |_{T\rightarrow
-6H^2}\right]/2}{1+\lambda\left[f_2(H)+12H^2\left .f_2'(T)\right
|_{T\rightarrow -6H^2}\right]}.
\ee
By substituting the matter density $\rho _m$ into Eq.~(\ref{H00}) we obtain 
the basic equation describing the cosmological dynamics in nonminimally
matter-coupled $f(T)$ gravity   as
\begin{widetext}
\begin{equation}\label{c2}
2\dot{H}=-\frac{\left\{1+\lambda  \left[f_2(H)+12 H^2 \left .f_2'(T)\right |_{T\rightarrow -6H^2}\right]\right\}
   \left\{\frac{ \left[f_1(H)+12 H^2 \left .f_1'(T)\right |_{T\rightarrow -6H^2}\right]/2+3
   H^2}{1+\lambda  \left[f_2(H)+12 H^2
   \left .f_2'(T)\right |_{T\rightarrow -6H^2}\right]}+p_m\right\}}{1+\left .f_1'(T)\right |_{T\rightarrow -6H^2}-12 H^2
   \left .f_1''(T)\right |_{T\rightarrow -6H^2}-\frac{2\lambda  \left\{ \left[f_1(H)+12 H^2
   \left .f_1'(T)\right |_{T\rightarrow -6H^2}\right]/2+3 H^2\right\} \left[\left .f_2'(T)\right |_{T\rightarrow -6H^2}-12
   H^2 \left .f_2''(T)\right |_{T\rightarrow -6H^2}\right]}{1+\lambda  \left[f_2(H)+12 H^2
   \left .f_2'(T)\right |_{T\rightarrow -6H^2}\right]}}.
   \end{equation}
   \end{widetext}

Once the functions $f_1(T)$ and $f_2(T)$ are fixed, Eqs.~(\ref{c1}) and
(\ref{c2}) become a system of two ordinary differential equations for three
unknowns, $\left(H,\rho _m,p_m\right)$. In order to close the system of
equations, the matter equation of state $p_m=p_m\left(\rho _m\right)$ must
also be given. Finally, the deceleration parameter is given by
\be
\label{deceleration}
q=-\frac{\dot{H}}{H^2}-1,
\ee
while the dark energy equation-of-state parameter can be
expressed as
\be
\!w_{DE}=-\frac{2\dot{H}}{\rho _{DE}}-\frac{\rho _m+p_m}{\rho
_{DE}}-1= -\frac{2\dot{H}+3H^2+p_m}{3H^2-\rho_m}.
\label{wDE2}
\ee

In the following we will investigate the system of Eqs.~(\ref{c1}) and
(\ref{c2}), for different functional forms of $f_1(T)$ and $f_2(T)$.

\subsection{$f_1(T)=-\Lambda+\alpha _1T^2$ and $f_2(T)=\beta _1T^2$}

As a first example, we examine the case where $f_1(T)=-\Lambda+\alpha _1T^2$
and $f_2(T)=\beta _1T^2$, where $\alpha _1$ and $\beta _1$ are
constants, since these are the first non-trivial corrections to TEGR, that
is to General Relativity. As we mentioned above, it proves convenient to
express the involved functions in terms of $H$. In particular, in
terms of $H$ the functional dependencies of $f_1$ and $f_2$ are given by
$f_1(H)=-\Lambda+\alpha H^4$ and $f_2(H)=\beta H^4$, respectively, with
$\alpha=36\alpha _1$, $\beta =36\beta _1$. For the derivatives of the
functions $f_1$ and $f_2$ we obtain $f_1'(H)=-\alpha H^2/3$, $f_2'(H)=-\beta
H^2/3$, $f_1''(H)=\alpha /18$, $f_2''(H)=\beta /18$. Moreover, we
restrict our analysis to the case of dust matter,  that is we take
$p_m=0$.

\begin{figure}
   \centering
  \includegraphics[width=8cm]{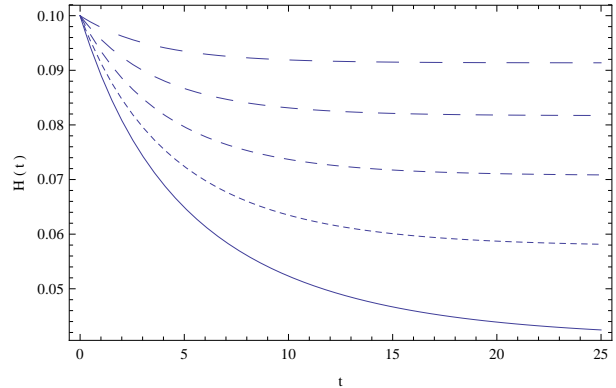}
  \caption{Time variation of the Hubble function $H(t)$ for the nonminimally
matter
coupled $f(T)$ gravity model with $f_1(T)=-\Lambda+\alpha _1T^2$ and
$f_2(T)=\beta _1T^2$, or equivalently   $f_1(H)=-\Lambda+\alpha H^4$ and
$f_2(H)=\beta H^4$ with $\alpha=36\alpha _1$, $\beta =36\beta _1$,
for five different choices of the parameters
$\Lambda$, $\alpha $, $\beta $, and $\lambda $: $\Lambda =0.01$, $\alpha
=0.16$, $\beta =0.1$, and $\lambda =1$ (solid curve), $\Lambda =0.02$,
$\alpha =0.18$, $\beta =0.3$, and $\lambda =1.2$ (dotted curve), $\Lambda
=0.03$, $\alpha =0.20$, $\beta =0.35$, and $\lambda =1.4$ (short-dashed
curve), $\Lambda =0.04$, $\alpha =0.30$, $\beta =0.45$, and $\lambda =1.6$
(dashed curve), and $\Lambda =0.05$, $\alpha =0.40$, $\beta =0.55$, and
$\lambda =1.8$ (long-dashed curve), respectively. The initial value for $H$
used to numerically integrate Eq.~(\ref{dem1}) is $H(0)=0.1$.} \label{fig1}
   \end{figure}

\begin{figure}[!]
   \centering
  \includegraphics[width=8cm]{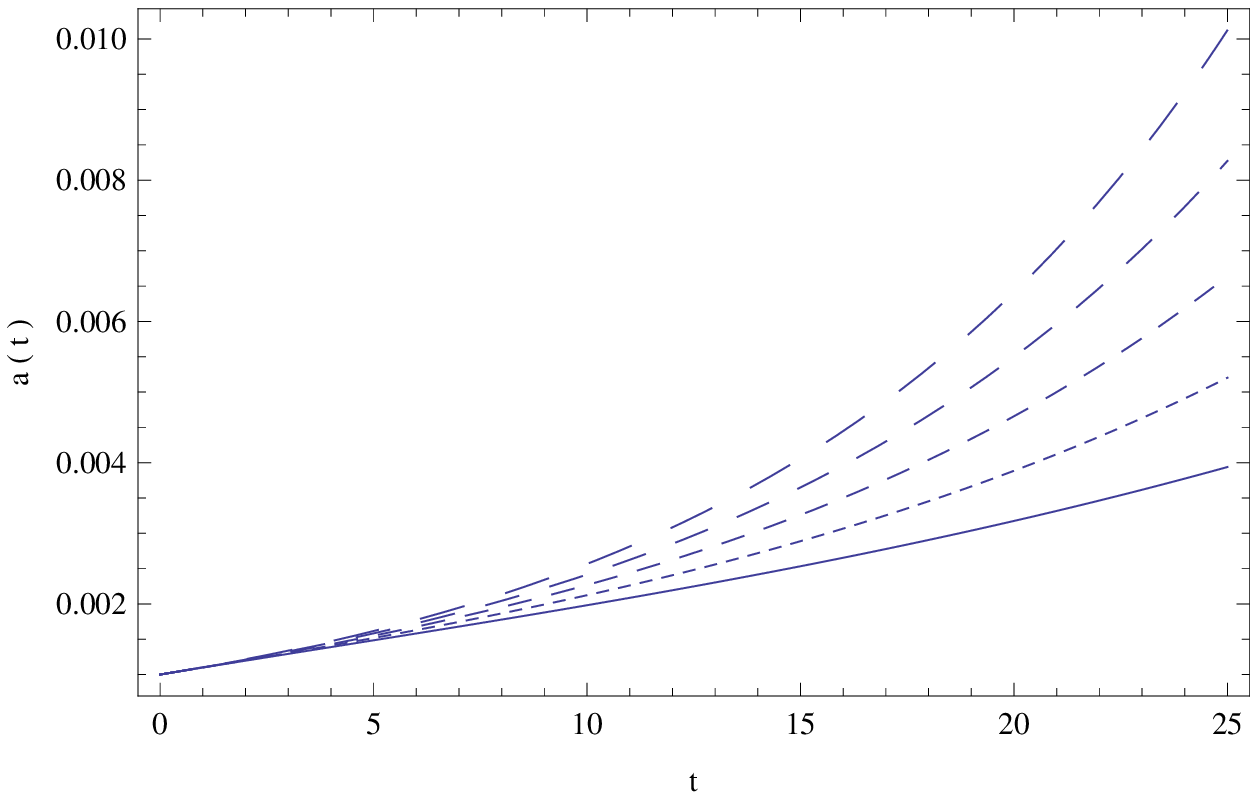}
  \caption{Time variation of of the scale factor $a(t)$ for the nonminimally
matter
coupled $f(T)$ gravity model with $f_1(T)=-\Lambda+\alpha _1T^2$ and
$f_2(T)=\beta _1T^2$, or equivalently   $f_1(H)=-\Lambda+\alpha H^4$ and
$f_2(H)=\beta H^4$ with $\alpha=36\alpha _1$, $\beta =36\beta _1$, for five
different choices of the parameters $\Lambda$, $\alpha $, $\beta $, and
$\lambda $: $\Lambda =0.01$, $\alpha =0.16$, $\beta =0.1$, and $\lambda =1$
(solid curve), $\Lambda =0.02$, $\alpha =0.18$, $\beta =0.3$, and $\lambda
=1.2$ (dotted curve), $\Lambda =0.03$, $\alpha =0.20$, $\beta =0.35$, and
$\lambda =1.4$ (short-dashed curve), $\Lambda =0.04$, $\alpha =0.30$, $\beta
=0.45$, and $\lambda =1.6$ (dashed curve), and $\Lambda =0.05$, $\alpha
=0.40$, $\beta =0.55$, and $\lambda =1.8$ (long-dashed curve), respectively.
The initial values for $a$ and $\dot{a}$ used to numerically integrate
Eq.~(\ref{dem1}) are $a(0)=10^{-3}$ and $\dot{a}(0)=10^{-4}$, respectively.}
\label{fig2}
   \end{figure}

\begin{figure}
   \centering
  \includegraphics[width=8cm]{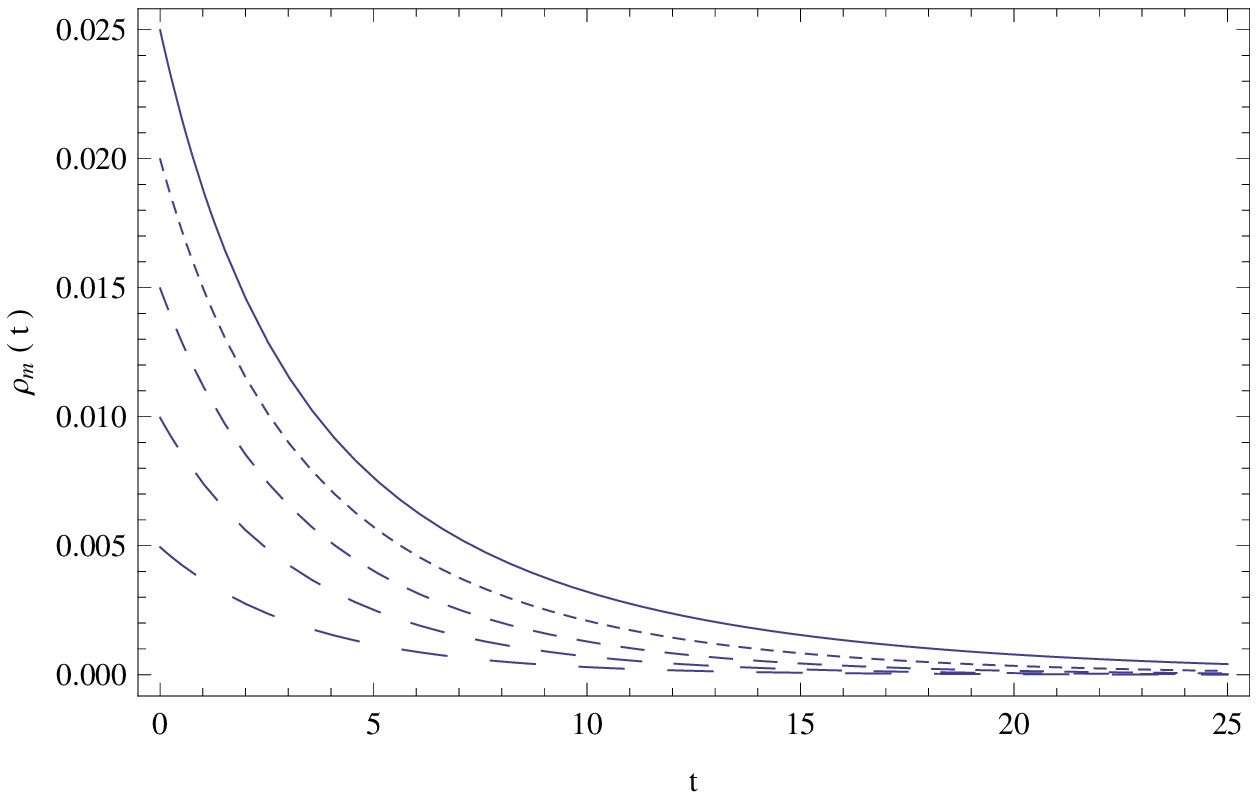}
  \caption{Time variation of the matter energy density $\rho _m(t)$ for the
nonminimally matter
coupled $f(T)$ gravity model with $f_1(T)=-\Lambda+\alpha _1T^2$ and
$f_2(T)=\beta _1T^2$, or equivalently   $f_1(H)=-\Lambda+\alpha H^4$ and
$f_2(H)=\beta H^4$ with $\alpha=36\alpha _1$, $\beta =36\beta _1$, for five
different choices of the parameters$\Lambda$, $\alpha $, $\beta $, and
$\lambda $: $\Lambda =0.01$, $\alpha =0.16$, $\beta =0.1$, and $\lambda =1$
(solid curve), $\Lambda =0.02$, $\alpha =0.18$, $\beta =0.3$, and $\lambda
=1.2$ (dotted curve), $\Lambda =0.03$, $\alpha =0.20$, $\beta =0.35$, and
$\lambda =1.4$ (short-dashed curve), $\Lambda =0.04$, $\alpha =0.30$, $\beta
=0.45$, and $\lambda =1.6$ (dashed curve), and $\Lambda =0.05$, $\alpha
=0.40$, $\beta =0.55$, and $\lambda =1.8$ (long-dashed curve), respectively.
The initial value for $H$ used to numerically integrate Eq.~(\ref{dem1}) is
$H(0)=0.1$.} \label{fig3}
   \end{figure}

\begin{figure}
   \centering
  \includegraphics[width=8cm]{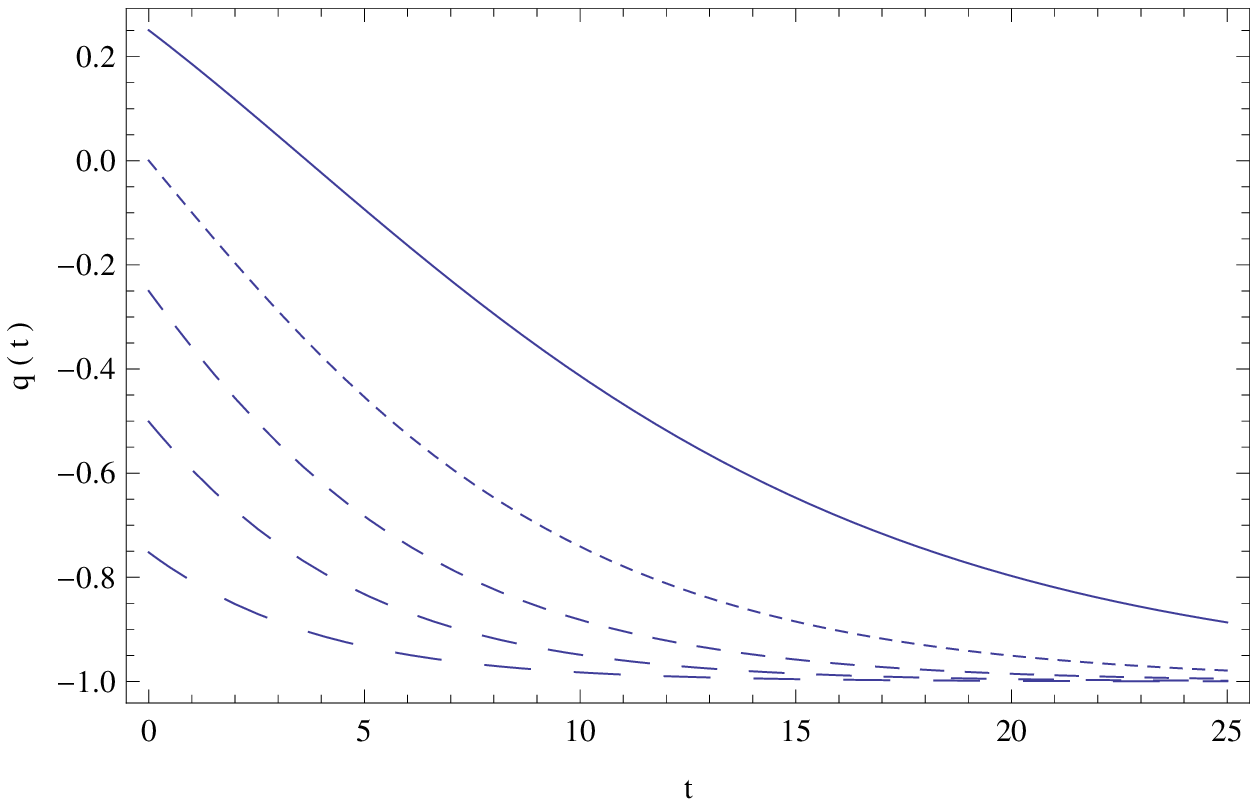}
  \caption{Time variation of the deceleration parameter $q(t)$ for
the nonminimally matter
coupled $f(T)$ gravity model with $f_1(T)=-\Lambda+\alpha _1T^2$ and
$f_2(T)=\beta _1T^2$, or equivalently   $f_1(H)=-\Lambda+\alpha H^4$ and
$f_2(H)=\beta H^4$ with $\alpha=36\alpha _1$, $\beta =36\beta _1$, for five
different choices of the parameters $\Lambda$, $\alpha $, $\beta $, and
$\lambda $: $\Lambda =0.01$, $\alpha =0.16$, $\beta =0.1$, and $\lambda =1$
(solid curve), $\Lambda =0.02$, $\alpha =0.18$, $\beta =0.3$, and $\lambda
=1.2$ (dotted curve), $\Lambda =0.03$, $\alpha =0.20$, $\beta =0.35$, and
$\lambda =1.4$ (short-dashed curve), $\Lambda =0.04$, $\alpha =0.30$, $\beta
=0.45$, and $\lambda =1.6$ (dashed curve), and $\Lambda =0.05$, $\alpha
=0.40$, $\beta =0.55$, and $\lambda =1.8$ (long-dashed curve), respectively.
The initial value for $H$ used to numerically integrate Eq.~(\ref{dem1}) is
$H(0)=0.1$.} \label{fig4}
   \end{figure}

\begin{figure}
   \centering
  \includegraphics[width=8cm]{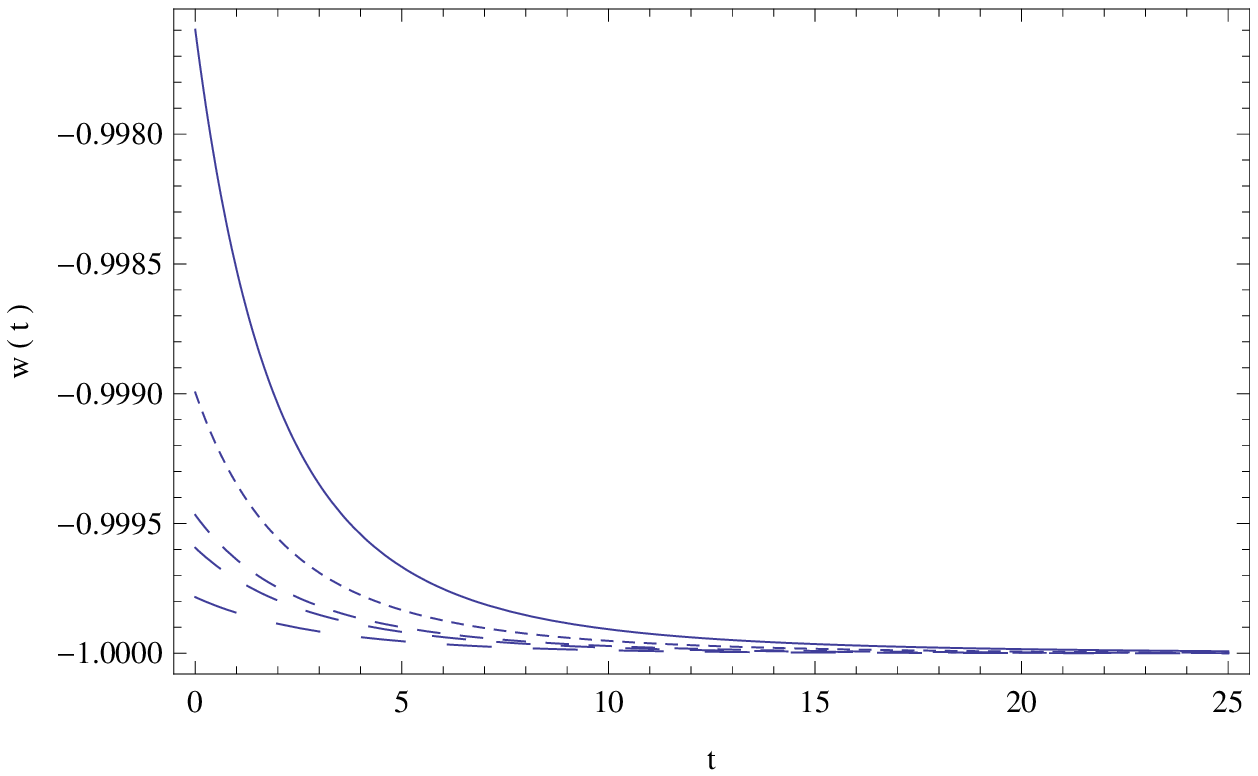}
  \caption{Time variation of the dark energy equation-of-state  parameter 
$w_{DE}$ for the nonminimally
matter
coupled $f(T)$ gravity model with $f_1(T)=-\Lambda+\alpha _1T^2$ and
$f_2(T)=\beta _1T^2$, or equivalently   $f_1(H)=-\Lambda+\alpha H^4$ and
$f_2(H)=\beta H^4$ with $\alpha=36\alpha _1$, $\beta =36\beta _1$,
for five different choices of the parameters
$\Lambda$, $\alpha $, $\beta $, and $\lambda $:
$\Lambda =0.01$, $\alpha =0.16$, $\beta =0.1$, and $\lambda =1$ (solid
curve), $\Lambda =0.02$, $\alpha =0.18$, $\beta =0.3$, and $\lambda =1.2$
(dotted curve), $\Lambda =0.03$, $\alpha =0.20$, $\beta =0.35$, and $\lambda
=1.4$ (short dashed curve), $\Lambda =0.04$, $\alpha =0.30$, $\beta =0.45$,
and $\lambda =1.6$ (dashed curve), and $\Lambda =0.05$, $\alpha =0.40$,
$\beta =0.55$, and $\lambda =1.8$ (long dashed curve), respectively. The
initial value for $H$ used to numerically integrate Eq.~(\ref{dem1}) is
$H(0)=0.1$.} \label{fig4a}
   \end{figure}

In this case the gravitational field equations
 (\ref{c1}) and (\ref{c2}) become
\be
\label{dem0}
\rho _m(t)=\frac{3 \alpha  H^4-6 H^2+\Lambda }{6 \beta  \lambda  H^4-2},
   \ee
   and
   \be\label{dem1}
   \dot{H}(t)=\frac{\left(3 \alpha  H^4-6 H^2+\Lambda \right) \left(3 \beta  \lambda
   H^4-1\right)}{4 H^2 \left(\alpha +\beta  \lambda  \Lambda -3 \beta  \lambda
    H^2\right)-4},
   \ee
   respectively. The time variation of the Hubble function $H$, of the scale 
factor $a$, of the matter energy density $\rho _m$, and of the deceleration
parameter $q$, obtained by numerically integrating Eqs.~(\ref{dem0}) and
(\ref{dem1}) are represented, for different numerical values of the free
parameters $\Lambda $, $\alpha $, $\beta $ and $\lambda $,  in
Figs.~\ref{fig1}-\ref{fig4a}.

As depicted in Fig.~\ref{fig1}, the Hubble function is a monotonically
decreasing function of time for all $t>0$. In the limit of large times the
Hubble function tends to a constant value, $\lim_{t\rightarrow
\infty}H(t)=h_0={\rm constant}$. Hence, for the considered range of values of
the free parameters, in the $f(T)$ model with torsion-matter coupling, the
Universe ends its evolution in an accelerating, de Sitter-type phase. The
scale factor $a$, shown in Fig.~\ref{fig2}, indicates a monotonically time
increase of the size of the Universe, and hence an expansionary behavior. The
matter energy density, depicted in Fig.~\ref{fig3}, tends progressively to
zero. Furthermore, the deceleration parameter $q$, presented in
Fig.~\ref{fig4}, indicates a large variety of dynamical behaviors of the
$f(T)$ model with matter-torsion coupling. In particular, for some values
of the free parameters the Universe starts its evolution in the
matter-dominated phase from a decelerating phase, and ends in a de
Sitter-type accelerated behavior. Other values of the parameters produce
Universe models starting from a marginally accelerating phase ($q=0$), and
ending in a de Sitter state. Finally, for other parameter choices
at the beginning of the matter dominated phase the Universe is
already in an accelerating phase, that is with $q<0$. Lastly, as depicted in
Fig.~\ref{fig4a}, for these specific choices of the parameters, the dark
energy equation-of-state parameter $w_{DE}$ is very close to the 
value $-1$, to which it rigorously tends in the large time limits. This is
an advantage, since in this model the effective torsion-matter coupling can
successfully mimic the cosmological constant, in agreement with observations.

After the above numerical elaboration, we examine whether we can obtain
analytical expressions in various limits. In particular, we analyze the
properties of the equations in the limit of small and large $H(t)$,
respectively.

\subsubsection{The limit of small $H$}

In the limit of small $H(t)$, that is at the late phases of the
cosmological evolution, Eq.~(\ref{dem1}) becomes
   \be
 \dot{H}=\frac{1}{4} \left(\alpha  \Lambda +\beta  \lambda  \Lambda
   ^2-6\right)H^2 +\frac{\Lambda }{4}
   \ee
   yielding the following solution
   \be 
H(t)=\sqrt{H_{1}}\tan \left[ \tan ^{-1}\left(
\frac{H_{0}}{\sqrt{H_{1}}}\right) +\frac{%
\Lambda }{4\sqrt{H_{1}}}\left( t-t_{0}\right) \right],
\label{Happrox1}
\ee
with $H_{1}=\Lambda /\left[\Lambda (\alpha +\beta \lambda \Lambda )-6%
\right]$, and  where we have used the initial condition
$H\left(t_0\right)=H_0$.
Thus, the scale factor evolves according to
   \be
a(t)=a_{0}\cos ^{-\frac{4H_{1}}{\Lambda }}\left[ \tan ^{-1}\left(
\frac{H_{0}%
}{\sqrt{H_{1}}}\right) +\frac{\Lambda }{4\sqrt{H_{1}}}\left( t-t_{0}\right) %
\right],
\ee
where $a_0$ is an arbitrary constant of integration.
  
Similarly, the deceleration parameter (\ref{deceleration}) becomes
   \be
   q(t)=-\frac{\Lambda  \csc ^2\left[\tan
^{-1}\left(\frac{H_0}{\sqrt{H_1}}\right)+\frac{\Lambda  \left(t-t_0\right)}{4
\sqrt{H_1}}\right]}{4 H_1}-1.
\ee
Additionally, the matter energy-density (\ref{dem0}) can be approximated as
  \be
   \rho _m(t)=3 H(t)^2-\frac{\Lambda }{2},
    \ee
   and using Eq. (\ref{Happrox1})
   its explicit time dependence
acquires the form
     \be
   \rho _m(t)=3 H_1 \tan ^2\left[\tan
^{-1}\left(\frac{H_0}{\sqrt{H_1}}\right)+\frac{\Lambda  \left(t-t_0\right)}{4
\sqrt{H_1}}\right]-\frac{\Lambda }{2}.
\ee


Finally, the dark-energy equation-of-state parameter from (\ref{wDE2}),
becomes
   \be
  w_{DE}(t)=-1-(\alpha+\beta\lambda \Lambda)H(t)^2,
   \ee
that is 
      \bea
w_{DE}(t)&=&-1-(\alpha+\beta\lambda
\Lambda)H_1  \times  \nonumber\\
&& \!\! \times  \tan ^2
\left[\tan
^{-1}\left(\frac{H_0}{\sqrt{H_1}}\right)+\frac{\Lambda  \left(t-t_0\right)}{4
\sqrt{H_1}}\right].
  \eea
Interestingly enough, we
observe that according to the parameter values, $w_{DE}$ can be either above
or below $-1$, that is the effective dark-energy sector can be
quintessence-like or phantom-like. This feature, which is expected to happen
in modified gravity \cite{Nojiri:2013ru}, is an additional advantage of the
scenario at hand.
   
Lastly, we mention that  in the large-time limit the Hubble function
(\ref{Happrox1}) becomes almost constant, implying that a de
Sitter-type evolution is possible in the framework of the present model.  

\subsubsection{The limit of large $H$}

In the limit of large $H$, corresponding to the early phases of the
cosmological evolution, in the first order approximation the differential
equation   (\ref{dem1})    describing the cosmological dynamics of
the Hubble function becomes
   \be
   \dot{H}=-\frac{3}{4} \alpha  H^4,
   \ee 
   with the general solution given by
   \be
   H(t)=\frac{2^{2/3}H_0}{\left[4+9H_0^3\alpha \left(t-t_0\right)\right]^{1/3}}.
   \ee
   The behavior of the scale factor can then be described by the equation
   \be
   a(t)=\exp\left\{\frac{1}{{3 \sqrt[3]{2}\alpha}}\left[\frac{4}{H_0^3}+9 \alpha  \left(t-t_0\right)\right]^{2/3}  \right\},
   \ee
   that is, it is determined solely by the parameter $\alpha$.
 Similarly, the deceleration parameter (\ref{deceleration})  is given by
   \be
   q(t)=\frac{3 \alpha H_0^2 }{\left[8+18 \alpha H_0^3 \left(t-t_0\right)\right]^{2/3}}-1.
   \ee
   Moreover, for the matter energy density (\ref{dem0})  we
obtain
   \be
   \rho _m(t)=\frac{\alpha }{2 \beta  \lambda },
   \ee
   showing that during the time interval for which this approximation is
valid the energy density of the matter is approximately constant.
Finally, the dark-energy equation-of-state parameter from (\ref{wDE2}),
becomes
   \be
  w_{DE}(t)=-\frac{ \alpha}{2} H(t)^2=-\frac{
\alpha}{2}
\frac{2^{4/3}H_0^2}{\left[4+9H_0^3\alpha
\left(t-t_0\right)\right]^{2/3}}.
   \ee
 Again, we mention that according to the parameter choice, $w_{DE}$ can be
either above or below $-1$, that is the effective dark-energy sector can be
quintessence-like or phantom-like.
   
\subsection{$f_1(T)=-\Lambda$ and $f_2(T)=\alpha _1T+\beta _1T^2$}

\begin{figure}[!]
   \centering
  \includegraphics[width=8cm]{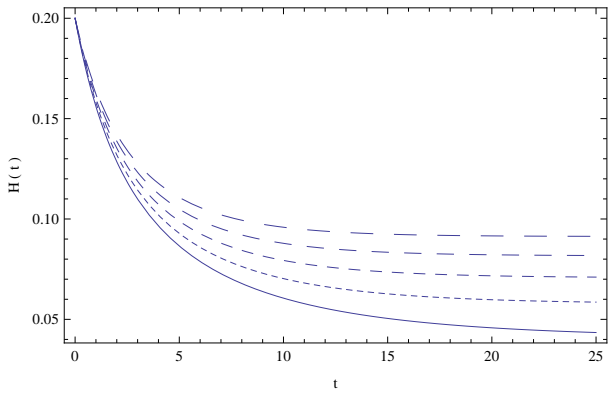}
  \caption{Time variation of the Hubble function $H(t)$ for the
nonminimally matter
coupled $f(T)$ gravity model with $f_1(T)=-\Lambda$ and $f_2(T)=\alpha
_1T+\beta _1T^2$, or equivalently  $f_1(H)=-\Lambda$ and $f_2(H)=\alpha
H^2+\beta H^4$ with $\alpha =-6\alpha_1$, $\beta =36\beta _1$,
for five different choices of the parameters $\Lambda$, $\alpha $, $\beta $,
and $\lambda $: $\Lambda =0.01$, $\alpha =0.16$, $\beta =0.1$, and $\lambda
=1$ (solid curve), $\Lambda =0.02$, $\alpha =0.18$, $\beta =0.3$, and
$\lambda =1.2$ (dotted curve), $\Lambda =0.03$, $\alpha =0.20$, $\beta
=0.35$, and $\lambda =1.4$ (short-dashed curve), $\Lambda =0.04$, $\alpha
=0.30$, $\beta =0.45$, and $\lambda =1.6$ (dashed curve), and $\Lambda
=0.05$, $\alpha =0.40$, $\beta =0.55$, and $\lambda =1.8$ (long-dashed
curve), respectively. The initial value for $H$ used to numerically integrate
Eq.~(\ref{mod22}) is $H(0)=0.2$.} \label{fig5}
   \end{figure}

\begin{figure}[!]
   \centering
  \includegraphics[width=8cm]{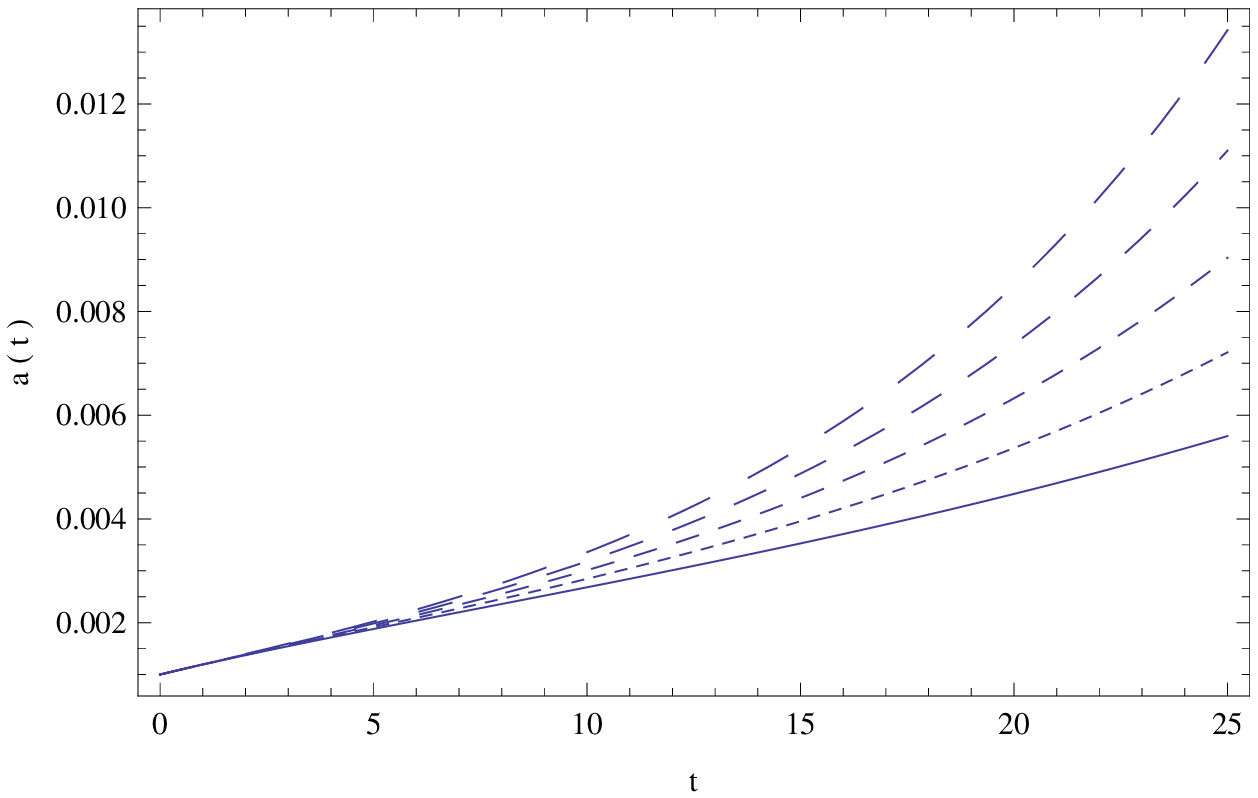}
  \caption{Time variation of the scale factor $a(t)$ of the Universe for the
nonminimally matter
coupled $f(T)$ gravity model with $f_1(T)=-\Lambda$ and $f_2(T)=\alpha
_1T+\beta _1T^2$, or equivalently  $f_1(H)=-\Lambda$ and $f_2(H)=\alpha
H^2+\beta H^4$ with $\alpha =-6\alpha_1$, $\beta =36\beta _1$,
for five different choices of the parameters $\Lambda$, $\alpha $, $\beta $,
and $\lambda $: $\Lambda =0.01$, $\alpha =0.16$, $\beta =0.1$, and $\lambda
=1$ (solid curve), $\Lambda =0.02$, $\alpha =0.18$, $\beta =0.3$, and
$\lambda =1.2$ (dotted curve), $\Lambda =0.03$, $\alpha =0.20$, $\beta
=0.35$, and $\lambda =1.4$ (short-dashed curve), $\Lambda =0.04$, $\alpha
=0.30$, $\beta =0.45$, and $\lambda =1.6$ (dashed curve), and $\Lambda
=0.05$, $\alpha =0.40$, $\beta =0.55$, and $\lambda =1.8$ (long-dashed
curve), respectively. The initial values for $a$ and $\dot{a}$ used to
numerically integrate Eq.~(\ref{mod22}) are $a(0)=10^{-3}$ and
$\dot{a}(0)=2\times 10^{-4}$, respectively.} \label{fig6}
   \end{figure}

\begin{figure}[!]
   \centering
  \includegraphics[width=8cm]{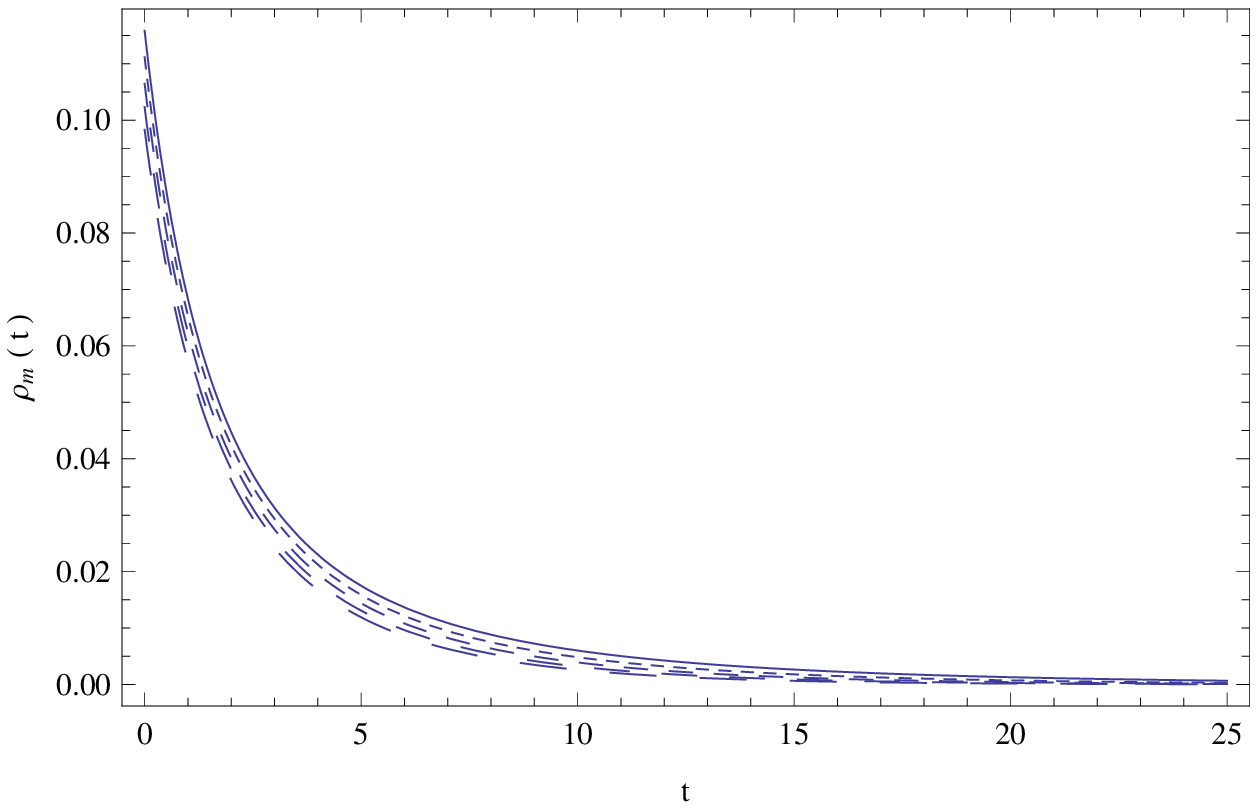}
  \caption{Time variation of the matter energy density $\rho _m(t)$ of the
Universe for the nonminimally matter
coupled $f(T)$ gravity model with $f_1(T)=-\Lambda$ and $f_2(T)=\alpha
_1T+\beta _1T^2$, or equivalently  $f_1(H)=-\Lambda$ and $f_2(H)=\alpha
H^2+\beta H^4$ with $\alpha =-6\alpha_1$, $\beta =36\beta _1$,
for five different choices of the parameters $\Lambda$, $\alpha $, $\beta $,
and $\lambda $: $\Lambda =0.01$, $\alpha =0.16$, $\beta =0.1$, and $\lambda
=1$ (solid curve), $\Lambda =0.02$, $\alpha =0.18$, $\beta =0.3$, and
$\lambda =1.2$ (dotted curve), $\Lambda =0.03$, $\alpha =0.20$, $\beta
=0.35$, and $\lambda =1.4$ (short dashed curve), $\Lambda =0.04$, $\alpha
=0.30$, $\beta =0.45$, and $\lambda =1.6$ (dashed curve), and $\Lambda
=0.05$, $\alpha =0.40$, $\beta =0.55$, and $\lambda =1.8$ (long dashed
curve), respectively. The initial value for $H$ used to numerically integrate
Eq.~(\ref{mod22}) is $H(0)=0.2$.} \label{fig7}
   \end{figure}

\begin{figure}[!]
   \centering
  \includegraphics[width=8cm]{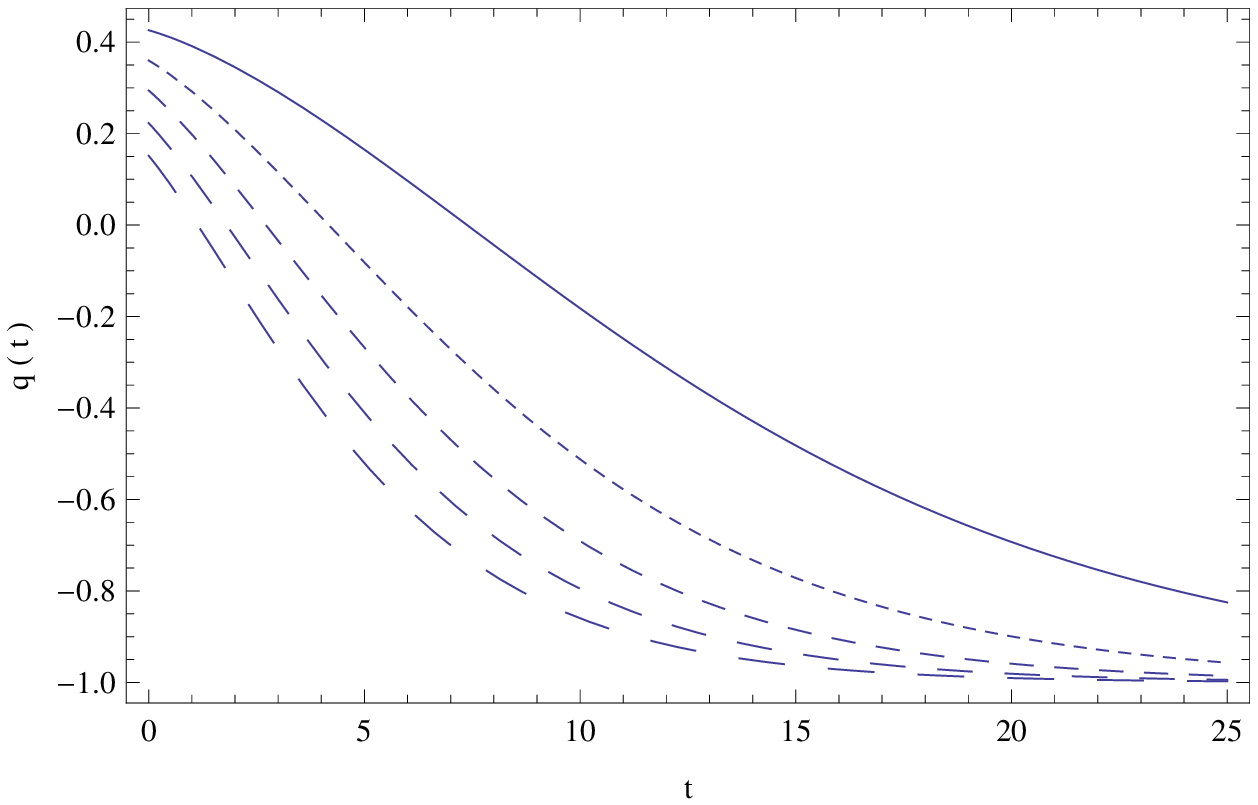}
  \caption{Time variation of the deceleration parameter $q(t)$ of the
Universe for the nonminimally matter
coupled $f(T)$ gravity model with $f_1(T)=-\Lambda$ and $f_2(T)=\alpha
_1T+\beta _1T^2$, or equivalently  $f_1(H)=-\Lambda$ and $f_2(H)=\alpha
H^2+\beta H^4$ with $\alpha =-6\alpha_1$, $\beta =36\beta _1$,
for five different choices of the parameters $\Lambda$, $\alpha $, $\beta $,
and $\lambda $: $\Lambda =0.01$, $\alpha =0.16$, $\beta =0.1$, and $\lambda
=1$ (solid curve), $\Lambda =0.02$, $\alpha =0.18$, $\beta =0.3$, and
$\lambda =1.2$ (dotted curve), $\Lambda =0.03$, $\alpha =0.20$, $\beta
=0.35$, and $\lambda =1.4$ (short dashed curve), $\Lambda =0.04$, $\alpha
=0.30$, $\beta =0.45$, and $\lambda =1.6$ (dashed curve), and $\Lambda
=0.05$, $\alpha =0.40$, $\beta =0.55$, and $\lambda =1.8$ (long dashed
curve), respectively. The initial value for $H$ used to numerically integrate
Eq.~(\ref{mod22}) is $H(0)=0.2$.}\label{fig8}
  \end{figure}

\begin{figure}
   \centering
  \includegraphics[width=8cm]{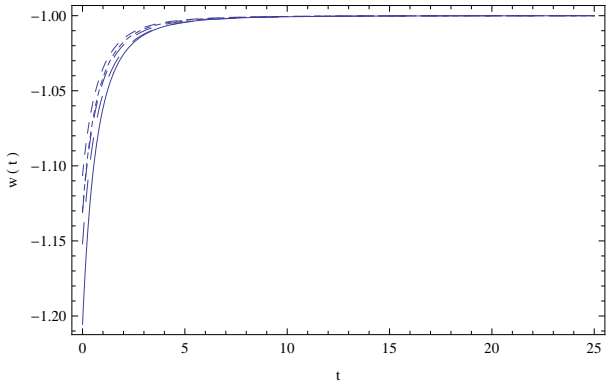}
  \caption{Time variation of the dark energy equation of state  parameter
$w_{DE}$ of the Universe  for the
nonminimally matter
coupled $f(T)$ gravity model with $f_1(T)=-\Lambda$ and $f_2(T)=\alpha
_1T+\beta _1T^2$, or equivalently  $f_1(H)=-\Lambda$ and $f_2(H)=\alpha
H^2+\beta H^4$ with $\alpha =-6\alpha_1$, $\beta =36\beta _1$,
for five different choices of the parameters $\Lambda$, $\alpha $, $\beta $,
and $\lambda $:
$\Lambda =0.01$, $\alpha =0.16$, $\beta =0.1$, and $\lambda =1$ (solid
curve), $\Lambda =0.02$, $\alpha =0.18$, $\beta =0.3$, and $\lambda =1.2$
(dotted curve), $\Lambda =0.03$, $\alpha =0.20$, $\beta =0.35$, and $\lambda
=1.4$ (short dashed curve), $\Lambda =0.04$, $\alpha =0.30$, $\beta =0.45$,
and $\lambda =1.6$ (dashed curve), and $\Lambda =0.05$, $\alpha =0.40$,
$\beta =0.55$, and $\lambda =1.8$ (long dashed curve), respectively. The
initial value for $H$ used to numerically integrate Eq.~(\ref{mod22}) is
$H(0)=0.2$.}\label{fig8a}
  \end{figure}

As a second example, we examine the case where $f_1(T)=-\Lambda$ and
$f_2(T)=\alpha _1T+\beta _1T^2$, where $\Lambda >0$, $\alpha _1$ and $\beta
_1$ are constants, since this scenario is also the first non-trivial
correction to TEGR, that is to General Relativity. Equivalently, we impose
$f_1(H)=-\Lambda$  and $f_2(H)=\alpha H^2+\beta  H^4$, with $\alpha =-6\alpha
_1$ and $\beta =36\beta _1$. For the derivatives of the functions $f_1(T)$
and
$f_2(T)$ we obtain $f_1'(T)=f_1''(T)=0$, $f_2'(H)=-\alpha /6-\beta
H^2/3$, and $f_2''(H)=\beta /18$.  The basic evolution equations  
 (\ref{c1}) and (\ref{c2}) in this case become 
\be
\label{mod21}
\rho _m(t)=\frac{\Lambda -6 H^2}{2 \alpha  \lambda  H^2+6 \beta  \lambda  H^4-2},
   \ee
   and
  \be\label{mod22}
   \dot{H}(t)=\frac{3 \left(\Lambda -6 H^2\right) \left(\alpha  \lambda  H^2+3 \beta
   \lambda  H^4-1\right)}{2 \left(\alpha  \lambda  \Lambda +6 \beta  \lambda
   H^2 \left(\Lambda -3 H^2\right)-6\right)},
   \ee
 respectively. The time variation of the Hubble function $H$, of the scale
factor $a$, of the matter density $\rho_m$, and of the deceleration
parameter, obtained by numerically elaborating the system of
Eqs.~(\ref{mod21}) and (\ref{mod22}) for different values of the free
parameters and assuming the matter to be dust ($w_m=0$),  are presented in
Figs.~\ref{fig5}-\ref{fig8a}, respectively.

The Hubble function, presented in Fig.~\ref{fig5}, decreases monotonically
in time, and tends to a constant value in the large-time limit. Therefore,
for all the parameter choices the Universe ends in a de Sitter
phase. The time variation of the scale factor, depicted in Fig.~(\ref{fig6}),
indicates that all considered models are expanding. The matter energy
density, shown in Fig.~\ref{fig7}, monotonically decreases in time
as expected. In the large-time limit the Universe ends in a vacuum state,
with negligible matter density, and thus being completely dominated by the
effective dark energy sector. The deceleration parameter, presented in
Fig.~\ref{fig8} indicates a very strong dependence of the dynamical behavior
of the Universe on the model parameters. For the considered values, in all
cases at the beginning of the matter-dominated phase, the Universe
is in a decelerating phase, with $q>0$. After a finite time $t_a$, determined
by the condition $q\left(t_a\right)=0$, the Universe enters in an accelerated
phase, with $q(t)<0,\forall t>t_a$. Similarly to the model of the previous
subsection, the Universe always ends in a de Sitter phase, with $q=-1$.
Finally, as depicted in Fig.~\ref{fig8a}, in the large time limit the dark
energy equation-of-state parameter $w_{DE}$ tends to the value $-1$, namely
$\lim_{t\rightarrow \infty }w_{DE}(t)=-1$, thus showing that this choice of
the functions $f_1(T)$ and $f_2(T)$ can also successfully mimic an effective
cosmological constant. Note however, that for these specific parameter
choices $w_{DE}$ lies in the phantom regime, which is an advantage of the
scenario at hand, revealing its capabilities.

After this numerical elaboration, we examine whether we can obtain
analytical expressions in various limits. In particular, we examine  the
properties of the equations in the limit of small and large $H(t)$,
respectively.

\subsubsection{The limit of small $H$}

In the limit of small values of the Hubble function $H$, that is at late
times,  Eq.~(\ref{mod22}) can be approximated
as
\be
\dot{H}=\frac{3 \Lambda }{2\left(6-\alpha  \lambda 
\Lambda\right)}+\frac{3\left(\alpha^2  \lambda^2 
\Lambda^2+6 \beta\lambda \Lambda^2-36\right)}{2\left(6-\alpha  \lambda 
\Lambda\right)^2}H^2,
\ee
with the general solution given by
 \be
H(t)=\sqrt{H_{2}}\tan \left[ \tan ^{-1}\left(
\frac{H_{0}}{\sqrt{H_{2}}}\right) +    \frac{3 \Lambda  \left(
t-t_{0}\right)}{2\left(6-\alpha \lambda 
\Lambda\right) \sqrt{H_{2}}  }      \right]  
\label{Happrox22}
\ee
with $H_2=\Lambda(6-\alpha\lambda\Lambda)/\left(\alpha^2  \lambda^2 
\Lambda^2+6 \beta\lambda \Lambda^2-36\right)$, and where we have used the
initial condition $H\left(t_0\right)=H_0$.
The scale factor then becomes
  \bea
a(t)&=&a_{0}\cos^{-\frac{2H_2(6-\alpha\lambda\Lambda)}{3\Lambda }} \times
   \nonumber \\
&& \!\! \times  \left[ \tan
^{-1}\left(
\frac{H_{0}}{\sqrt{H_{2}}}\right) +    \frac{3 \Lambda  \left(
t-t_{0}\right)}{2\left(6-\alpha \lambda 
\Lambda\right) \sqrt{H_{2}}  }      \right]     ,
\eea
where $a_0$ is an arbitrary constant of integration, while the deceleration
parameter (\ref{deceleration}) behaves as
\be
q(t)=-\frac{3\Lambda  \csc ^2\left[ \tan
^{-1}\left(
\frac{H_{0}}{\sqrt{H_{2}}}\right) +    \frac{3 \Lambda  \left(
t-t_{0}\right)}{2\left(6-\alpha \lambda 
\Lambda\right) \sqrt{H_{2}}  }      \right]   }{2\left(6-\alpha \lambda 
\Lambda\right) H_2}-1.
\ee
Note that in the limit of large $t$ the deceleration parameter tends to
$-1$, $\lim _{t\rightarrow \infty}q(t)=-1$. Additionally, the matter density
(\ref{mod21}) becomes
\be
   \rho _m(t)=\left(3-\frac{\alpha\lambda\Lambda}{2}\right)
H(t)^2-\frac{\Lambda }{2},
    \ee
   and using (\ref{Happrox22}) we acquire
\bea
 \rho _m(t)&=&  -\frac{\Lambda}{2}+\left(3-\frac{\alpha\lambda\Lambda}{2}\right) H_2 \times
  \nonumber\\
 && \hspace{-1.0cm} \times \tan^2 \left[ \tan ^{-1}\left(
\frac{H_{0}}{\sqrt{H_{2}}}\right) +    \frac{3 \Lambda  \left(
t-t_{0}\right)}{2\left(6-\alpha \lambda 
\Lambda\right) \sqrt{H_{2}}  }      \right] .
\eea
Finally, the dark-energy equation-of-state parameter from (\ref{wDE2}),
becomes
{\small{
  \bea
   w_{DE}(t)&=&-1+\frac{\alpha\lambda\Lambda}{
\alpha\lambda\Lambda-6}  -\frac {
18\left(\alpha^2\lambda^2\Lambda+2\beta\lambda\Lambda-6\alpha\lambda
\right)}{ \left(6-\alpha\lambda\Lambda\right)^2}H^2
    \nonumber\\
&=&-1+\frac{\alpha\lambda\Lambda}{
\alpha\lambda\Lambda-6}  -\frac {
18\left(\alpha^2\lambda^2\Lambda+2\beta\lambda\Lambda-6\alpha\lambda
\right)}{ \left(6-\alpha\lambda\Lambda\right)^2}H_2 
     \nonumber\\
 &&\times   \tan^2 \left[ \tan ^{-1}\left(
\frac{H_{0}}{\sqrt{H_{2}}}\right) +    \frac{3 \Lambda  \left(
t-t_{0}\right)}{2\left(6-\alpha \lambda 
\Lambda\right) \sqrt{H_{2}}  }      \right] .
   \eea}}
which can lie both in the quintessence as well as in the phantom regime, 
depending on the specific choices of the free parameters of the model,
namely on $\alpha $, $\beta $, $\lambda $ and $\Lambda $, respectively.

\subsubsection{The limit of large $H$}

In the opposite limit of large $H$, that is at early times, at first
order approximation Eq.~(\ref{mod22}) becomes
\be
\dot{H}-\frac{3}{2} H^2=0,
\ee
with $H\left(t_0\right)=H_0$, and thus the general solution is given by
\be
H(t)= \frac{2 H_0}{2-3 H_0 \left(t-t_0\right)}.
\ee
 The scale factor then reads
\be
a(t)=a_{0} \left[\frac
{2}{2-3 H_{0} \left(t-t_0\right)}\right]^{2/3},
   \ee
while the deceleration parameter is obtained as $q=-5/2$, that is the
universe at early times always starts with acceleration, which corresponds
to an inflationary stage. Finally, for the time variation of the matter
energy density in the large-$H$ regime we find $\rho_m(t)\approx
0$, which is consistent with the interpretation of this stage as inflationary.

We mention that the above expressions for $H$, $a$, $q$ and $\rho_m$ at
first order approximation, are independent on the free parameters of the
model $\alpha $, $\beta $, $\lambda $ and $\Lambda $, respectively, and are
determined only by the initial value of $H$ at $t=t_0$.

\section{Discussions and final remarks}\label{Concl}

In the present paper, we have considered an extension of the $f(T)$ gravity
model by introducing a nonminimal coupling between torsion and matter. The
geometric part of the action was extended through the introduction of two
independent functions of the torsion scalar $T$, namely $f_1(T)$ and
$f_2(T)$, respectively, with the function $f_2(T)$ being nonminimally
coupled to the matter Lagrangian $L_m$. The resulting gravitational model
presents some formal analogies with the nonminimally geometry-matter
coupling introduced in \cite{Bertolami:2007gv}. However, the resulting
equations, as well as its physical and geometrical interpretations, are
very different. The theory of nonminimal torsion-matter coupling is
therefore a novel class of gravitational modification. 

From the physical point of view, in this theory,  matter is not just a
passive component in the space-time continuum, but it plays an active role in
the overall gravitational dynamics, which is strongly modified due to the
supplementary interaction between matter and geometry. Moreover, the major
advantage of the $f(T)$-type models, namely that the field equations are
second order, is not modified by the torsion-matter coupling. 

As an application of the nonminimal torsion-matter coupling scenario we
have considered the dynamical evolution of a flat FRW universe. We
have investigated the time dependence of the cosmologically relevant physical
parameters, for two different choices of the functions $f_1(T)$ and $f_2(T)$,
corresponding to the simplest departures from General Relativity.
In these specific models the dynamics of the Universe
is determined by the free parameters which appear in the functions $f_1(T)$
and $f_2(T)$,  as well as by the matter-torsion coupling constant. Depending
on the numerical values of these parameters a large number of cosmological
behaviors can be obtained. In our analysis we have considered the matter
dominated phase of the Universe evolution, that is, we neglected the matter
pressure. More general models with $p_m$ can be easily constructed and
analyzed.

We restricted our analysis in expanding evolutions, although contracting or 
bouncing 
solutions can be easily obtained as well. We have found a universe evolution
in agreement with observations, that is a matter-dominated era followed by
an accelerating phase. Additionally, the effective dark-energy
equation-of-state parameter can lie in the quintessence or phantom regime,
which reveals the capabilities of the scenario. Furthermore, a general and
common property of the considered models is that they all end in a de Sitter
phase, with zero matter density, that is to complete dark-energy domination.
Finally, these models also accept solutions with almost constant Hubble
function, which can describe the inflationary regime. Thus, the scenario of
nonminimal torsion-matter coupling can offer a unified description of the
universe evolution, from its inflationary to the late-time accelerated
phases.

Apart from the exact numerical elaboration, we have extracted approximate
analytical expressions in the limit of a small Hubble parameter, that is
corresponding to the large-time limit, as well as for large Hubble
parameters, that is corresponding to the beginning of the cosmological
expansion. These expressions verify the above physical features that were
extracted through the numerical analysis.

In conclusion, based on the torsional formulation of gravity, we have
proposed a novel modified gravitational scenario which contains an
arbitrary coupling between the torsion scalar and the matter Lagrangian. The
cosmological implications of this theory proves to be very interesting. 
However, in order for the present scenario to be considered as a
good candidate for the description of Nature, additional investigations
should be performed, such as the detailed comparison with cosmological
observations, the complete perturbation analysis, etc. These necessary
studies lie beyond the scope of the present work and are left for future
projects.

\begin{acknowledgments} 
FSNL acknowledges financial  support of the Funda\c{c}\~{a}o para a
Ci\^{e}ncia e Tecnologia through an Investigador FCT Research contract, with
reference IF/00859/2012, funded by FCT/MCTES (Portugal), and grants
CERN/FP/123615/2011, CERN/FP/123618/2011 and EXPL/FIS-AST/1608/2013. GO would like to thank CAPES 
and FAPEMIG for financial support. The research of ENS is implemented within the
framework of the Operational Program ``Education and Lifelong Learning''
(Actions Beneficiary: General Secretariat for Research and Technology), and
is co-financed by the European Social Fund (ESF) and the Greek State. 
\end{acknowledgments}

\end{document}